\documentclass[aps, prb, twocolumn, showpacs, superscriptaddress, preprintnumbers, amsmath, amssymb]{revtex4-1}
\usepackage[latin1]{inputenc}
\usepackage[T1]{fontenc}
\usepackage{graphicx}
\usepackage{dcolumn}
\usepackage{bm}
\usepackage{color}
\usepackage[pdfborder={0 0 0}, colorlinks=true,citecolor=blue,linkcolor=black,urlcolor=blue]{hyperref}

\begin{document}

\title{Crystal growth and the electronic phase diagram of the 4$d$ doped Na$_{1-\delta}$Fe$_{1-x}$Rh$_x$As in comparison with 3$d$ doped Na$_{1-\delta}$Fe$_{1-x}$Co$_x$As}
\author{Frank Steckel}
\email{f.steckel@ifw-dresden.de}
\affiliation{Leibniz Institute for Solid State and Materials Research, 01171 Dresden, Germany}
\author{Maria Roslova}
\thanks{F. Steckel and M. Roslova contributed equally to this work}
\affiliation{Leibniz Institute for Solid State and Materials Research, 01171 Dresden, Germany}
\affiliation{Department of Chemistry, Lomonosov Moscow State University, 119991 Moscow, Russia}
\author{Robert Beck}
\affiliation{Leibniz Institute for Solid State and Materials Research, 01171 Dresden, Germany}
\author{Igor Morozov}
\affiliation{Leibniz Institute for Solid State and Materials Research, 01171 Dresden, Germany}
\affiliation{Department of Chemistry, Lomonosov Moscow State University, 119991 Moscow, Russia}
\author{Saicharan Aswartham}
\affiliation{Leibniz Institute for Solid State and Materials Research, 01171 Dresden, Germany}
\affiliation{Deptartment of Physics and Astronomy, University of Kentucky, Lexington, Kentucky 40506-0055 USA}
\author{Daniil Evtushinsky}
\author{Christian G. F. Blum}
\affiliation{Leibniz Institute for Solid State and Materials Research, 01171 Dresden, Germany}
\author{Mahmoud Abdel-Hafiez}
\affiliation{Leibniz Institute for Solid State and Materials Research, 01171 Dresden, Germany}
\affiliation{Faculty of Science, Physics Department, Fayoum University, 63514 Fayoum, Egypt}
\author{Dirk Bombor}
\author{Janek Maletz}
\author{Sergey Borisenko}
\affiliation{Leibniz Institute for Solid State and Materials Research, 01171 Dresden, Germany}
\author{Andrei V. Shevelkov}
\affiliation{Department of Chemistry, Lomonosov Moscow State University, 119991 Moscow, Russia}
\author{Anja U. B. Wolter}
\affiliation{Leibniz Institute for Solid State and Materials Research, 01171 Dresden, Germany}
\author{Christian Hess}
\affiliation{Leibniz Institute for Solid State and Materials Research, 01171 Dresden, Germany}
\affiliation{Center for Transport and Devices, Technische Universit\"at Dresden, 01069 Dresden, Germany}
\author{Sabine Wurmehl}
\email{s.wurmehl@ifw-dresden.de}
\affiliation{Leibniz Institute for Solid State and Materials Research, 01171 Dresden, Germany}
\affiliation{Institut für Festkörperphysik, TU Dresden, 01062 Dresden, Germany}
\author{Bernd B\"uchner}
\affiliation{Leibniz Institute for Solid State and Materials Research, 01171 Dresden, Germany}
\affiliation{Institut für Festkörperphysik, TU Dresden, 01062 Dresden, Germany}
\affiliation{Center for Transport and Devices, Technische Universit\"at Dresden, 01069 Dresden, Germany}

\date{\today}

\begin{abstract}
Single crystals of Na$_{1-\delta}$Fe$_{1-x}$T$_x$As with T = Co, Rh have been grown using a self-flux technique. The crystals were thoroughly characterized by powder X-ray diffraction, magnetic susceptibility and electronic transport with particular focus on the Rh-doped samples. Measurements of the specific heat and ARPES were conducted exemplarily for the optimally doped compositions. The spin-density wave transition (SDW) observed for samples with low Rh concentration ($0\,\leq\,x\,\leq\,0.013$) is 
fully suppressed in the optimally doped sample. The superconducting transition temperature ($T_c$) is enhanced from $10$~K in Na$_{1-\delta}$FeAs to $21$~K in the optimally doped sample ($x$ = 0.019) of the Na$_{1-\delta}$Fe$_{1-x}$Rh$_x$As series and decreases 
for the overdoped compounds, revealing a typical shape for the superconducting part of the electronic phase diagram. Remarkably, the phase diagram is almost identical to that of Co-doped Na$_{1-\delta}$FeAs, suggesting a generic phase diagram for both dopants.
%

\end{abstract}



\maketitle

\section{Introduction}
Iron-based superconductors typically exhibit a canonical emergence of the superconducting phase from an orthorhombically distorted spin-density wave (SDW) parent state where the latter is suppressed in favour of superconductivity upon doping, or by pressure. Well known examples are the so-called 1111 and 122 families of compounds\cite{Cruz2008, Huang2008, Luetkens2009, Zhi-An2008, Rotter2008, Chu2009a, Hess2009, Ni2009, Chu2009, Colombier2009}. Remarkably, for the most extensively studied material  Ba(Fe$_{1-x}$T$_x$)$_2$As$_2$ (T~=~transition metal) it has been shown that primarily the number of $d$-electrons of the dopant is decisive for the actual doping evolution of the electronic phases. More specifically, for T~=~Co, Rh and T~=~Ni, Pd, which pairwise are located in the same group of the periodic table, the respective electronic phase diagrams are essentially identical\cite{Ni2009}.

The physics of the layered iron arsenide Na$_{1-\delta}$FeAs resembles in many aspects that of the 1111 and 122 phases described above. Yet, there are important differences. It exhibits an orthorhombically distorted SDW phase with structural and magnetic transitions at 50~K and 40~K, respectively,\cite{Parker2009,Parker2010,Wright2012,Presniakov2013} i.e. at an about 100~K lower transition temperature as is observed for its undoped 1111 and 122 pendants\cite{Klauss2008,Luetkens2009,Rotter2008,Rotter2008a}. Furthermore, superconductivity is observed to coexist with SDW order below $\sim10$~K, however, with a very small superconducting volume fraction\cite{Parker2009,Parker2010,Wright2012,Chen2009,Sasmal2009}. A very detailed electronic phase diagram has been established as a function of Co content which substitutes Fe, i.e. for Na$_{1-\delta}$Fe$_{1-x}$Co$_x$As.\cite{Parker2010,Wright2012,Wang2012} Like in the 1111 and 122 compounds,\cite{Sefat2008,Prando2013,Ni2009} this kind of doping yields a suppression of the SDW and bulk superconductivity, but the maximum critical temperature $T_c\sim 20$~K is achieved at a much lower Co concentration compared to the BaFe$_2$As$_2$ series. 

The impact of other dopants in Na$_{1-\delta}$FeAs different from Co is scarcely explored. Bulk superconductivity at selected doping levels has been reported for Ni- and Pd-doped Na$_{1-\delta}$FeAs \cite{Parker2010,Steckel2014a}. Furthermore, Cu substitution for Fe has been reported to induce bulk superconductivity at doping levels between 0.5\% and $\sim3.3$\%.\cite{Wang2013}

Na$_{1-\delta}$FeAs crystalizes in the anti-PbFCl structure type (space group $P4/nmm$)\cite{Parker2009,Johrendt2011,Stewart2011} and possesses a more rigid crystal structure than e.g. the 122 compounds. This provides less space for various substitutions on the cation as well as on the anion sublattices. One may, thus, expect that doping with 4$d$ metals may induce considerable structural changes due to significantly larger atomic radii of the respective dopants. Such 
changes may include variations of Fe-As bond lengths and distortions of FeAs$_4$ tetrahedral motifs, leading to significant alterations in the electronic structure. Thus, it is important to investigate the substitution of Fe  by 4$d$ elements in Na$_{1-\delta}$FeAs and to compare the resulting structure and physical properties of 3$d$ and 4$d$ element doped Na$_{1-\delta}$Fe$_{1-x}$T$_x$As.

In this work, we present the superconducting and normal-state properties of single crystalline Na$_{1-\delta}$Fe$_{1-x}$Rh$_x$As studied by means of magnetic susceptibility, and, for selected compositions, electrical resistivity and specific heat measurements. For one selected single crystal, namely the optimally doped one, ARPES measurements are presented to monitor the superconducting transition and to underline the good quality of our crystal. The comparison of the electronic phase diagrams of the Na$_{1-\delta}$Fe$_{1-x}$T$_x$As series with 3$d$ and 4$d$ dopants (T=Rh, Co) shows that the Co-substituted samples (3$d$) as well as the Rh-substituted samples (4$d$) cause a similar change of critical temperatures as a function of increasing dopant content with a well-defined maximum value around 21~K.

\section{Experimental Details}

\begin{figure}
\centering
\includegraphics[width=8cm]{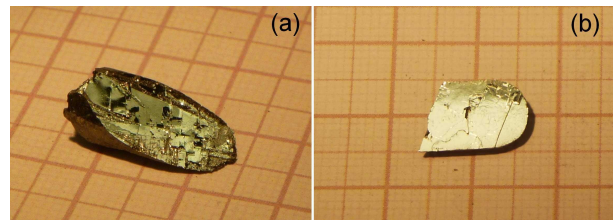}
\caption{(Color online) Freshly cleaved crystals of (a) Na$_{1-\delta}$Fe$_{0.99}$Co$_{0.01}$As, (b) Na$_{1-\delta}$Fe$_{0.981}$Rh$_{0.019}$As on scale paper to 
illustrate their size and to show the metallic shiny appearance of their cleavage surface.}
\label{fig-01}
\end{figure}

Single crystals of Na$_{1-\delta}$FeAs and Na$_{1-\delta}$Fe$_{1-x}$T$_x$As with different contents of T = Co and Rh were grown by a self-flux technique. All preparation steps were performed in an argon-filled glove box with an O$_2$ and H$_2$O content of less than 0.1~ppm. The Na-As precursor used for the synthesis was obtained from Na lumps (99.95\%, Alfa Aesar) and As powder (99.99\%, Chempur). The stoichiometric Na:As mixture was sealed in a niobium crucible and heated up first to 300°C and then up to 600°C, kept at each temperature for 5~h and after that was cooled rapidly down to room temperature. Two different strategies for the crystal growth were applied: (i) prereacted Na-As, metallic Fe (99.998\%, Puratronic) and T, where T~=~Co (99.8\%, Heraeus) or T~=~Rh (99.9\%, Saxonia), in a molar ratio of Na-As:(Fe+T)~=~2.3:1; (ii) lumps of metallic Na and powders of elemental Fe, T and As were used for the growth of 4$d$ element doped Na$_{1-\delta}$FeAs crystals. In both cases, the total amount of 6~g of material was placed in an alumina crucible inside a niobium container which was welded under 1~atm of Ar in an arc-melting facility. The niobium container was sealed in an evacuated quartz tube afterwards and heated slowly up to 950°C (1150°C for T~=~Rh), kept at this temperature for 15~h, and cooled down to 600°C with a rate of 3°C/h. Thin plate-like single crystals with a maximum size of about 10$\times$10$\times$0.05~mm$^3$ were extracted mechanically from the ingot.
All crystals grew in a layered morphology; they can be easily cleaved along the $ab$ plane. Fig.~\ref{fig-01} exemplarily shows two crystals of the Na$_{1-\delta}$Fe$_{1-x}$T$_x$As series, one where Fe is substituted by Co, the other with Rh substitution on the Fe site, to illustrate the size and the metallic, shiny appearance of the crystals. Note, that the crystals are extremely air sensitive, hence, all manipulations during the sample preparation for the physical measurements were performed in an argon-filled glovebox.

\begin{figure}
\centering
\includegraphics[width=8cm]{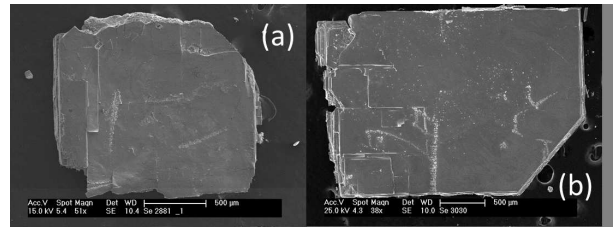}
\caption{SEM images of single crystalline (a) Na$_{1-\delta}$Fe$_{0.978}$Co$_{0.023}$As, (b) Na$_{1-\delta}$Fe$_{0.987}$Rh$_{0.013}$As,
showing their layered growth morphology.}
\label{fig-02}
\end{figure}

The microstructure and composition of the single crystals were analyzed by scanning electron microscopy (SEM, XL30 Philipps, IN400) equipped with an electron microprobe analyzer for semi-quantitative elemental analysis using the energy dispersive mode (EDX). Fig.~\ref{fig-02} shows SEM images of freshly cleaved surfaces of two Na$_{1-\delta}$Fe$_{1-x}$T$_x$As (T~=~Co, Rh) samples. The plate-like morphology of these crystals reflects the layer-by-layer growth of these compounds. The composition of all crystals is determined by taking EDX spectra and averaging the data over several points of the same specimen and for several crystals of each batch. Although the Na K$_{\alpha}$-line at 1.04~keV is clearly pronounced in the EDX spectra, the error of the specific Na content by this technique may be large (see discussion below). In the following, we refer to specific samples by their respective dopant content as measured by EDX. X-ray powder diffraction data were collected on a Stoe Stadi P diffractometer in transmission geometry with MoK$_{\alpha1}$-radiation equipped with a Ge monochromator and a DECTRIS MYTHEN 1~K detector. The samples were protected by sealing inside a glass capillary to prevent degradation in air. The powder diffraction pattern was scanned over the angular range \mbox{5~-~49°} with a step size of $\Delta(2\theta)=0.01$°. Profile analysis including LeBail refinement was done using the JANA2006 software.

Magnetization measurements were performed using either a Quantum Design Magnetic Properties Measurement System (MPMS 5T) or a Vibrating Sample
Magnetometer (SQUID-VSM 7T) with the field ($H=20$~Oe) applied parallel to the $ab$ plane of the crystals in a temperature range 1.8 - 30~K under zero-field 
and field-cooling conditions. For the investigation of the magnetization in the temperature range 1.8 - 300~K, the field ($H=1$~T) was applied
parallel to the $ab$ plane of the crystals after cooling in zero field.\\

Resistivity measurements were performed under cryogenic vacuum in the temperature range \mbox{4.3 - 300~K} in a home-made device using a standard four-probe technique. Electrical contacts parallel to the $ab$ plane were made using thin copper wires attached to the sample surface with a silver epoxy. All procedures including the contacts preparation and crystal mounting onto the probe were performed in an argon-filled glove box which ensured constantly inert conditions for the sample.\\ 

Specific heat measurements were carried out at a Quantum Design 9~T Physical Properties Measurement System (PPMS). During the heat capacity
measurements, the sample was cooled to the lowest temperature in an applied magnetic field (fc) and the specific heat data were obtained between 
1.8~K and 60~K (upon warming) using the relaxation time method. \\

Angle-resolved photoemission spectroscopy (ARPES) measurements were carried out at the $1^3$ end station at the UE-112pgm beamline of the BESSY-II synchrotron (Helmholtz-Zentrum Berlin f\"ur Materialien und Energie). Data were collected from a freshly cleaved smooth sample surface.

\section{Results and Discussion}

\subsection{Composition determination and crystallographic structure}
\label{xrd}
\begin{figure}
\centering
\includegraphics[width=8.5cm]{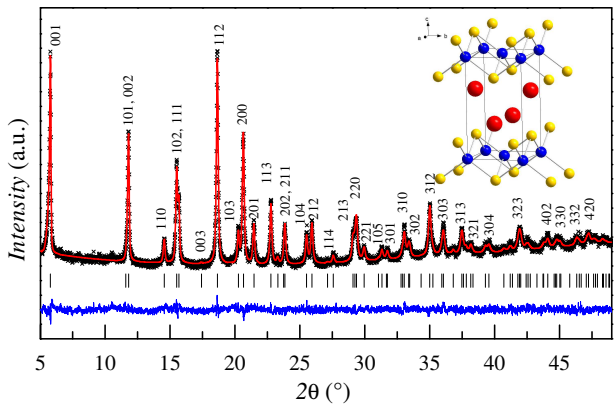}
\caption{(Color online) Powder x-ray diffraction pattern for Na$_{1-\delta}$Fe$_{0.957}$Rh$_{0.043}$As taken with MoK$_{\alpha1}$-radiation. The LeBail refinement is plotted as a solid line, with the Miller indices notated. The difference of the refinement and the measurement is indicated by the lower solid line. The structure is shown as an inset.}
\label{fig-03}
\end{figure}

Based on the EDX data the Na content for all the Na$_{1-\delta}$Fe$_{1-x}$T$_x$As samples under investigation was found to be $(1-\delta)=0.9$. Note that the composition data are normalized to $\Sigma(\mathrm{Fe+T})=1$ based on the assumption of a full occupancy of the Fe site and that the Na content may be underestimated by EDX in the presence of the higher atomic number atoms Fe and As.

A typical powder diffraction pattern for a Na$_{1-\delta}$Fe$_{1-x}$Rh$_x$As crystal with $x=0.043$ is given in Fig.~\ref{fig-03}. In order to monitor the evolution of lattice parameters, we applied a simple LeBail refinement to extract the very basic crystallographic information on Na$_{1-\delta}$Fe$_{1-x}$T$_x$As. The peak shape was assumed to be a pseudo-Voigt function and the refinement included the following aspects: (i) the background, which was modeled by a Legendre polynomial function with 14 terms; (ii) the scale factors; (iii) the global instrumental parameters (zero-point 2$\theta$ shift and systematic shifts, depending on transparency and off-centering of the sample); (iv) the lattice parameters and (v) the profile parameters (Caglioti half-width parameters of the pseudo-Voigt function). The texture correction was included using the March-Dollase function. For some crystals, we found minor amounts (<5w\%) of the foreign phases NaOH and FeAs due to decomposition of the samples and of Na$_3$As and Na$_3$As$_7$ from remaining flux. Since those foreign phases may in some cases not be well-crystallized, they do not appear as sharp reflections in all cases but are sometimes manifested as a broad hump at low diffraction angles. The refined lattice parameters for Na$_{1-\delta}$Fe$_{1-x}$T$_x$As with $\mathrm{T}=\mathrm{Co}$ and Rh are given in Table~\ref{tab:table-1}.

\begin{figure}
\centering
\includegraphics[width=8.5cm]{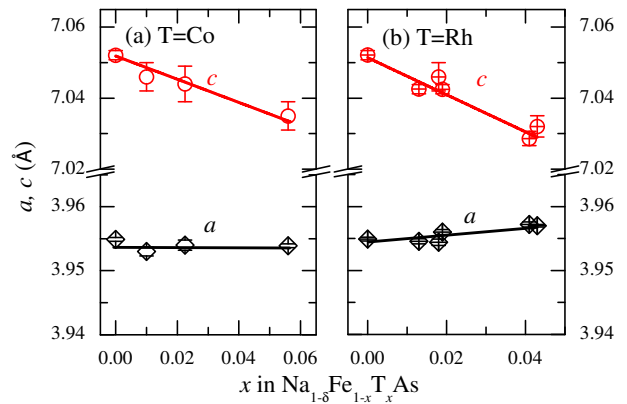}
\caption{(Color online) Comparison of the $a$ and $c$ lattice parameters upon doping for (a) the Na$_{1-\delta}$Fe$_{1-x}$Co$_x$As and (b) the Na$_{1-\delta}$Fe$_{1-x}$Rh$_x$As series.}
\label{fig-04}
\end{figure}


The transition metal composition was inferred from the EDX data, the resulting Fe to Rh, Co ratio was fixed and not further refined. In order to estimate the influence of the doping level for 4$d$ substituted samples on the $a$ and $c$ lattice parameters we compared the Na$_{1-\delta}$Fe$_{1-x}$T$_x$As series with T~=~Rh and Co. The results of the X-ray powder diffraction analysis are summarized in Fig.~\ref{fig-04}.

As expected, the lattice parameter $c$ decreases with increasing Co- and Rh-doping content; the measured lattice parameters for the Co-doped compounds are comparable to recent publications\cite{Parker2010,Wang2012}. Since there are a few reports in literature on series of Co-doped samples, we can compare our results with respect to published lattice constants: The ratio between the $c$ lattice constants of Co-doped and undoped Na$_{1-\delta}$FeAs: $c(x)/c(0)$ is 99.7\% for the $x=0.056$ compound, which represents the highest doping level in our study. This ratio is somewhat larger than that previously reported by Wang \textit{et al.} \cite{Wang2012}, but agrees well with result of Parker \textit{et al.} \cite{Parker2010}. A linear regression of the evolution of the $c$ lattice parameter yields a slope of -0.3 for the Co-doped compounds and a slope of -0.52 for the Rh-doped compounds. The ratio between slopes of roughly a factor of 1.7 reflects the larger radius of Rh in comparison with the radius of Co.

\begin{table*}[ht]
	\centering
	\begin{tabular}{| c | c | c | c | c | c | c |}
	\hline
	Dopant & $x$ & \multicolumn{2}{c|}{Lattice parameters} & \multicolumn{3}{c|}{transition temperatures}\\
	\hline
	\rule{0pt}{10pt} & & $a$ (\r{A}) & $c$ (\r{A}) & $T_{S}$~(K) & $T_{SDW}$~(K) & $T_c$~(K) \\
	\hline
	\hline
	Na$_{1-\delta}$Fe$_{1-x}$T$_x$As & 0 & 3.9549(3) & 7.0521(13) & 52 & 41.5 & 10 \\
	\hline
	 & 0.010 & 3.952(6) & 7.046(1) & 47 & 35 & 13.6\\
	 T~=~Co & 0.023 & 3.953(3) & 7.044(5) & & & 21.4  \\
	 & 0.056 & 3.950(4) & 7.035(7) & & & 15.7\\
	\hline
	 & 0.013 & 3.9545(4) & 7.0425(13) & 36 & 25.5 & 14.9  \\
	 & 0.018 & 3.954(4) & 7.046(6) &  &  & 21.9 \\
	 T~=~Rh & 0.019 & 3.956(5) & 7.039(4)&&& 22.0 \\		
	 & 0.041 & 3.957(2) & 7.031(3) &&& 17.8\\		
	 & 0.043 & 3.957(4) & 7.028(2) &&& 18.1\\		
	\hline
	\end{tabular}
 \caption{Lattice parameters of Na$_{1-\delta}$Fe$_{1-x}$T$_x$As with T~=~Co, Rh (space group P4/nmm, $T=298$~K and phase transition temperatures from resistivity and magnetization measurements).}
	\label{tab:table-1}	
\end{table*}

\subsection{Magnetization}
\label{mag}

\begin{figure}
\centering
\includegraphics[width=8.5cm]{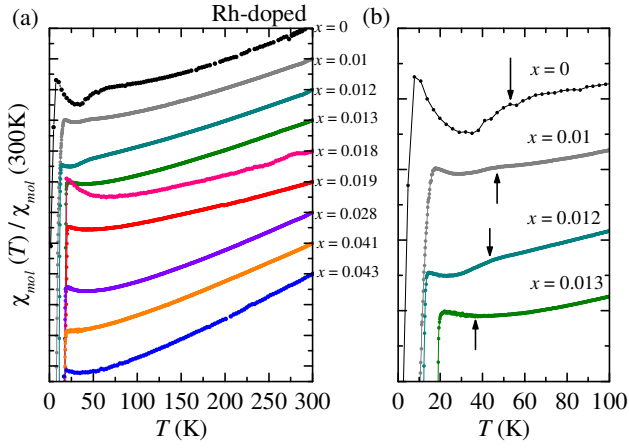}
\caption{(Color online) Normalized and shifted normal state susceptibility curves in an applied field of 1~T of Na$_{1-\delta}$Fe$_{1-x}$Rh$_x$As for different Rh contents. In (b) only the region of the SDW and structural phase transitions is shown for the parent compound and doping levels up to 0.013 of Rh. The structural phase transition temperatures are marked with an arrow.}
\label{fig-05a}
\end{figure}
\begin{figure}
\centering
\includegraphics[width=8.5cm]{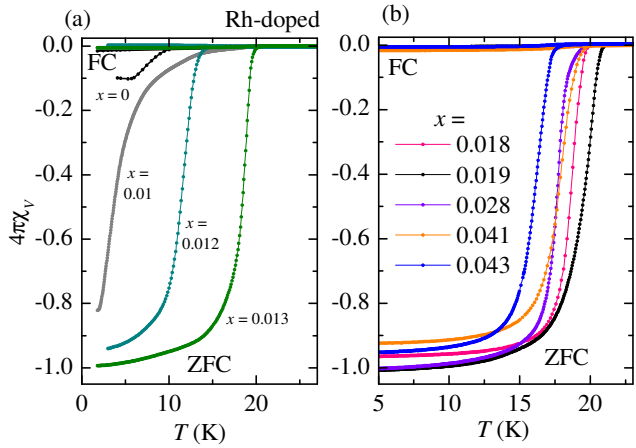}
\caption{(Color online) Zero-field and field-cooled susceptibility curves at an applied field of 20~Oe. (a) Underdoped Na$_{1-\delta}$Fe$_{1-x}$Rh$_x$As
($x$ = 0, 0.01, 0.012, 0.013). (b) Optimally and overdoped Na$_{1-\delta}$Fe$_{1-x}$Rh$_x$As ($x$ = 0.018, 0.019, 0.028, 0.041, 0.043). $T_c$ increases upon doping until $x=0.019$.}
\label{fig-05}
\end{figure}

\begin{figure}
\centering
\includegraphics[width=8.5cm]{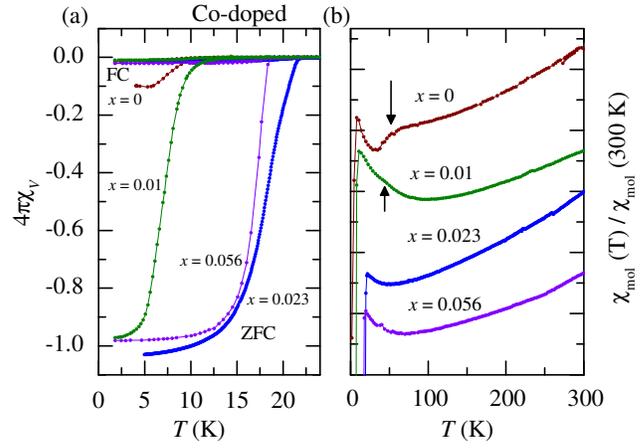}
\caption{(Color online) (a) Zero-field and field-cooled susceptibility curves in an applied field of 20~Oe of Na$_{1-\delta}$Fe$_{1-x}$Co$_x$As 
($x$ = 0, 0.01, 0.023, 0.056). $T_c$ increases upon doping until 0.02 of Co, simultaneously the volume fraction increases to $100$\%,
(b) Normalized normal-state susceptibility curves at an applied field of $1$~T of Na$_{1-\delta}$Fe$_{1-x}$Co$_x$As: the structural phase transition anomalies are marked with an arrow.}
\label{fig-06}
\end{figure}

Fig.~\ref{fig-05a} presents the normal-state in-plane magnetic susceptibility in an applied field of $H=1$~T for Na$_{1-\delta}$Fe$_{1-x}$T$_x$As with T~=~Rh. We compare them with selected T~=~Co data (Fig.~\ref{fig-06}~(b)) which agree with similar previous reports \cite{Parker2010, Wang2012, Wright2012}. The rapid drop associated with the superconducting transition at low temperatures is observed for all Rh-doped samples. For the Rh-doped samples with $x=0.01$ and $x=0.012$ pronounced features on the $1$~T curve associated with the structural transition are clearly visible. For the $x=0.013$ Rh-doped sample the same anomaly is hardly visible, but nevertheless recognizable and clearly resolved in transport, allowing a meaningful analysis. The transition temperatures are marked in Fig.~\ref{fig-05a}(b). Thus, $x=0.01$ up to $x=0.013$ are located on the underdoped side of the superconducting dome.

The normal-state susceptibility data demonstrate a positive slope with an almost linear behaviour above 100~K for all Rh-doping contents. Such a linear dependence of the high-temperature normal-state susceptibility was previously observed for other families of iron-based superconductors including 
LaFeAsO$_{1-x}$F$_x$\cite{Klingeler2010}, SrFe$_2$As$_2$\cite{Yan2008}, Ca(Fe$_{1-x}$Co$_x$)$_2$As$_2$\cite{Klingeler2010}, Ba$_{1-x}$Na$_x$Fe$_2$As$_2$\cite{Aswartham2012} and Ba(Fe$_{1-x}$T$_x$)$_2$As$_2$\cite{Ni2010}. Such a linear behavior of the magnetic susceptibility has been found in different calculations for 2D Fermi gases\cite{Korshunov2009} or LDA-DMFT\cite{Skornyakov2011,Skornyakov2012} or for antiferromagnetic fluctuations of local SDW correlations\cite{Zhang2009}.

Fig.~\ref{fig-05} and \ref{fig-06}~(a) show the volume susceptibility taken under zero-field (ZFC) and field-cooled (FC) conditions with $H=20$~Oe applied parallel to the $ab$ plane of Na$_{1-\delta}$Fe$_{1-x}$T$_x$As (T~=~Co, Rh) single crystals. As reported previously\cite{Parker2010,Wang2012,Wright2012} the Na$_{1-\delta}$FeAs parent compound with $T_c=10$~K shows only a very small diamagnetic response, along with a superconducting volume fraction of $\sim10$\%. The absence of bulk superconductivity is confirmed by the lack of an anomaly at $T_c$ in the specific heat (see section D.~Specific heat).

Doping with Rh as well as with Co strongly increases the value of the superconducting shielding fraction. Bulk superconductivity is achieved at the doping level $x=0.01$ in the NaFe$_{1-x}$Co$_x$As samples series and $x=0.012$ in the NaFe$_{1-x}$Rh$_x$As series. $T_c$ is obtained from the magnetization as the bifurcation temperature of the ZFC and FC curves. The critical temperature $T_c$ is enhanced from $10$~K for the Na$_{1-\delta}$FeAs parent compound to $21.2$~K in the optimally Rh-doped sample ($x\sim0.019$). The superconducting transition temperature $T_c$ decreases down to $18.3$~K for the Rh content of $x=0.043$, viz. for the overdoped side of the superconducting regime. It would be interesting to investigate the physical properties also at higher Rh doping levels but due to the poor solubility of Rh in the NaAs flux it was not possible to grow crystals with a Rh content higher than $x=0.043$.

\subsection{Transport}
\label{tr}

\begin{figure}
\centering
\includegraphics[width=8.5cm]{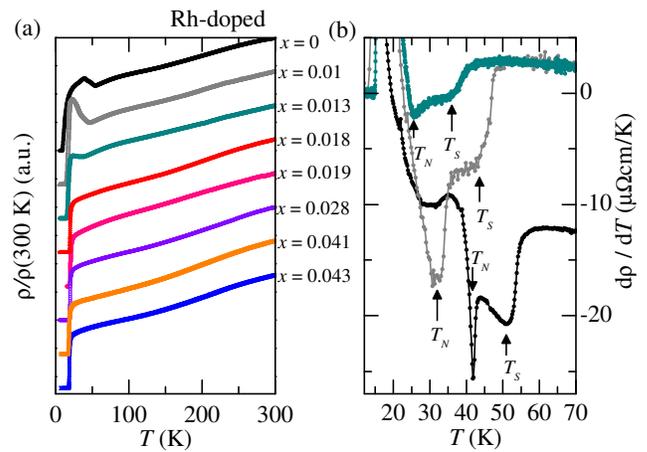}
\caption{(Color online) (a) Normalized and shifted resistivity of Na$_{1-\delta}$Fe$_{1-x}$Rh$_x$As for $x$~=~0, 0.01, 0.013, 0.018, 0.019, 0.028, 0.042, 0.043 in the full range of 4.3 to 300~K. (b) First derivative of the resistivity of the underdoped samples. The derivation of the undoped crystal is shifted by -15~$\mu\Omega$cm/K for better visibility. Anomalies due to the structural and magnetic phase transitions are marked with an arrow.}
\label{fig-07a}
\end{figure}

Fig.~\ref{fig-07a} presents the resistivity measured on Na$_{1-\delta}$Fe$_{1-x}$Rh$_x$As crystals. The resistivity at room temperature can be specified to be in a wide range of $0.2-1.2$~m$\Omega$cm. This unusual spread in the absolute values might be caused by the geometric error but more likely it
is due to cracks affecting the effective geometric factor. Such enhanced effective geometric factor was equally found on Co-doped NaFeAs\cite{Spyrison2012}.

The high temperature behavior ($300$~K~$>T>50$~K) seems to be nearly unchanged upon doping up to a concentration of $x=0.013$ pointing to a small effect of the Rh-doping on the charge carrier scattering\cite{Spyrison2012}.

Typical anomalies associated with the structural ($T_S$) and spin-density wave ($T_{SDW}$) transitions are clearly seen in the derivative of the resistivity for the undoped Na$_{1-\delta}$FeAs compound at around 52~K and 42~K, which is in good agreement with the resistivity data given in earlier work\cite{Parker2010,Spyrison2012,Wang2012,Wright2012,Tanatar2012} and in line with our magnetization and specific heat data. Upon Rh-doping these anomalies shift to lower temperatures and change slightly their appearance (anomalies are marked with an arrow in Fig.~\ref{fig-07a}~(b)). At the doping level of $x=0.01$ the SDW phase transition is indicated by a change of slope of the resistivity at 32~K. At lower temperature the resistivity increases further and peaks due to the superconducting transition. Despite the lowering of the phase transition temperatures the connected transport anomalies are less pronounced. Already at $x=0.018$ the formerly seen anomalies corresponding to the structural and magnetic phase transitions are completely suppressed. Concomitantly with this suppression of the structural and magnetic phase transitions, the superconducting temperature $T_c$ increases strongly by more than a factor of two. $T_c$  is determined by the temperature at which $\rho$ reaches zero resistance. The maximum $T_c$ is found to be very close to 
the one observed for optimally doped Na$_{1-\delta}$Fe$_{1-x}$Co$_x$As (21~K). Both Co- and Rh-doped samples with a dopant content $x\sim0.02$ show similar $T_c$ values. 


Together with the suppression of the structural and magnetic phases with increasing Rh doping the high temperature resistivity seems to slightly change in the intermediate range between $70$~K~$<T<200$~K. A very broad dip, similarly mentioned by Tanatar \textit{et al}.\cite{Tanatar2012}, is visible in the normal-state temperature region. The change of slope is typically a sign of changed charge carrier scattering. The scattering could be enhanced by fluctuations in the vicinity of the structural transition like in LaFeAsO\cite{Hess2009}. The broad dip, which we interpret as a sign of these fluctuations is still visible even in the overdoped samples. This would point to a very broad and extended fluctuation regime in the normal state. On the other hand these changes in the slope of the resistivity can be explained by a change in the charge carrier density or by an effective mass renormalization, too.\cite{Spyrison2012} 



\subsection{Specific heat}
\label{sh}

\begin{figure}
 \includegraphics[width=8.5cm]{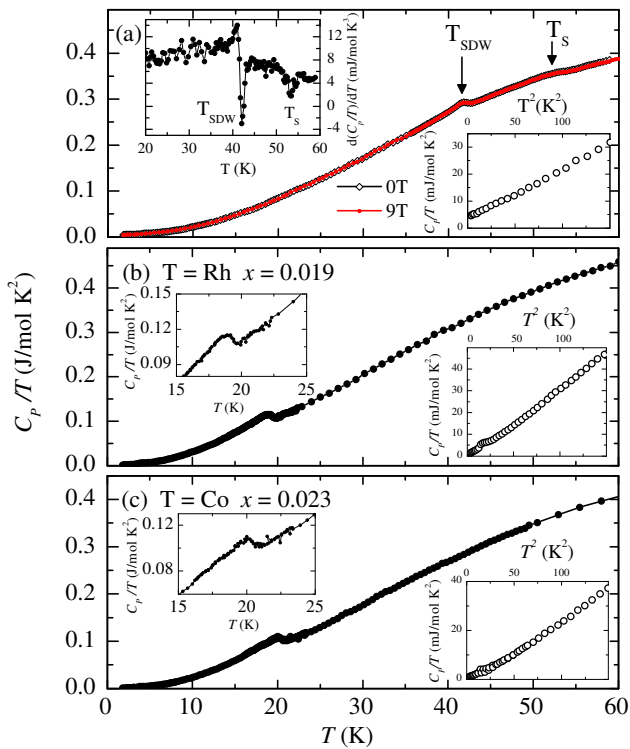}
\caption{(Color online) (a) Temperature dependence of the specific
heat in a Na$_{1-\delta}$FeAs single crystal. The zero-field and
9~T measurements show no typical anomaly for superconductivity at
low temperatures, while the structural and magnetic phase transition
anomalies are observed. The upper inset shows the derivative of
the specific heat around the phase transitions. (b) and (c)
present the temperature-dependent specific heat of
Na$_{1-\delta}$Fe$_{1-x}$T$_{x}$As single crystals with T~=~Rh and
Co and $x=0.019$ and 0.023, respectively. The lower insets display
the low-temperature region with a linear behavior in $C_p/T$ vs.
$T^{2}$. The upper insets are a zoom to the critical temperature
region.} 
 \label{fig:1}
 \end{figure}

Specific heat measurements on Na$_{1-\delta}$Fe$_{1-x}$T$_x$As
single crystals with $x=0$, $x=0.019$ (T~=~Rh) and $x=0.023$ (T~=~Co), were performed to further investigate the thermodynamic properties of these compounds (see Fig.~\ref{fig:1}~(a), (b), and (c) respectively). Fig.~\ref{fig:1} displays the temperature dependence of the specific heat (plotted as $C_{p}/T$) in the temperature range of 2-60~K. Our specific heat data for the parent compound Na$_{1-\delta}$FeAs shown in Fig.~\ref{fig:1}~(a) match nicely with that of previous reports\cite{Chen2009, Wang2012}. Two distinct features found at 51.7~K and 41.6~K correspond to the bulk structural and SDW transitions. The upper inset of Fig.~\ref{fig:1}~(a) presents the derivative of the specific heat, where the structural and SDW transitions can be inferred from the narrow minima. Please note that the observation of narrow features highlights the quality of our crystals. The corresponding temperatures found by specific heat are in good agreement with the values found by magnetization and transport as well as with data reported in literature\cite{Chen2009,Wang2012}. No jump is observed in the specific heat at low temperatures around 10~K, which can be explained by a very small superconducting volume fraction in the Na$_{1-\delta}$FeAs parent compound. Both the temperature dependence of the specific heat in zero field and in 9~T are very similar. This similar behavior has also been observed in EuFe$_{2}$As$_{2}$, where the sharp specific heat jump around 185~K is not affected even in fields of $H =14$~T\cite{Wu2009}.

In Figs.~\ref{fig:1}(b) and (c) a clear sharp anomaly is observed in the specific heat measurements proving the bulk nature of superconductivity in these samples and yielding $T_c$ values of 19.4~K and 20.9~K for Na$_{1-\delta}$Fe$_{1-x}$T$_x$As with $T$ = Rh and Co, respectively. The critical temperatures as derived from the specific heat data are in agreement with the $T_c$ found by resistivity and magnetization measurements. The lower insets of Fig.~\ref{fig:1} represent the temperature dependence of the low temperature part of the specific heat data plotted as $C_p/T$ vs. $T^{2}$. In the Rh- and the Co-doped crystal we found a tiny entropy contribution at 15~K$^2$ which might be a hint towards an impurity contribution, but the size of the anomaly is too small for a proper analysis. 

\subsection{ARPES}
\begin{figure}
\includegraphics[width=8.5cm]{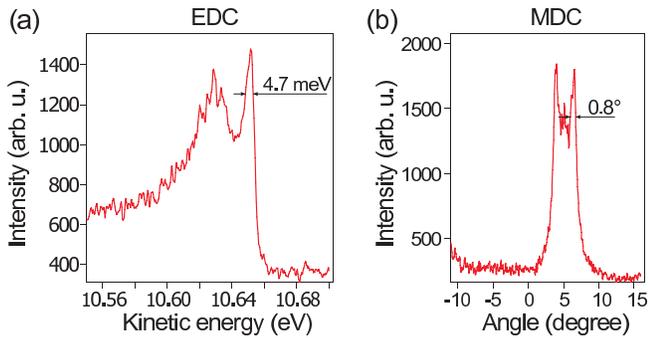}
\caption{(Color online) (a) The width of the energy distribution curve of Na$_{1-\delta}$Fe$_{1-x}$Rh$_x$As ($x=0.019$) is less than 5~meV at 15~K. (b) The width of the momentum distribution curve is less than 1 degree.}
\label{fig-ARPES2}
\end{figure}
\begin{figure}
\includegraphics[width=6.5cm]{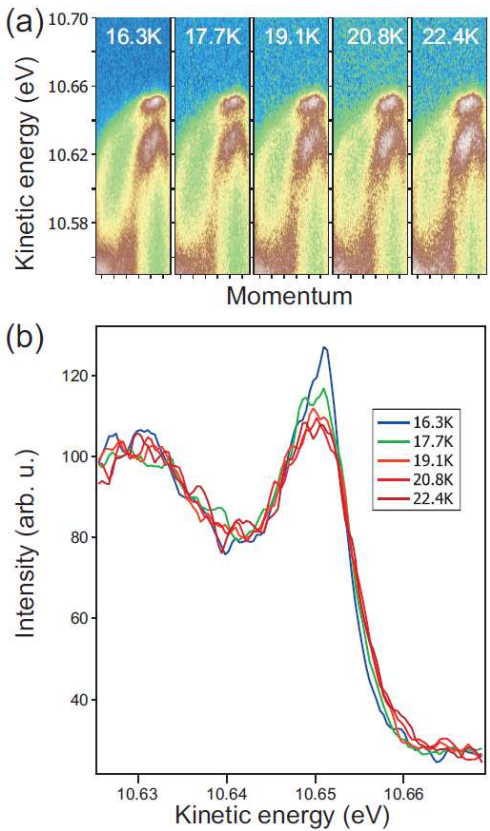}
\caption{(Color online) (a) Energy-momentum cut passing through the hole-like electronic bands at the center of the Brillouine zone, recorded at 16.3, 17.7, 19.1,
20.8 and 22.4~K in Na$_{1-\delta}$Fe$_{1-x}$Rh$_x$As with $x=0.019$. A transition between 17.7 and 19.1~K can be seen in the intensity distribution. (b) The temperature dependence of the energy distribution curve reveals a growth of the peak at the Fermi level and a shift of the leading edge of the spectrum, evidencing a superconducting transition at 18-19~K.}
\label{fig-ARPES}
\end{figure}

Generally, the Fermi surface (FS) of lightly electron-doped NaFeAs is formed by two electron-like FS sheets at the corner of the Brillouin zone (BZ), the M point, and three hole-like bands, barely touching the Fermi level at the BZ center, the $\Gamma$ point.\cite{Thirupathaiah2012} Both experimental studies and theoretical calculations show high suitability of 111 iron arsenides for ARPES measurements\cite{Borisenko2010,Lankau2010}, largely due to the presence of a natural cleavage plane between two layers of alkali metal atoms.
Consequently, unlike the situation for some other iron-based superconductors, one does not expect any hindrances to the observation of the superconducting transition in ARPES experiments on
111 systems\cite{Borisenko2010,Thirupathaiah2012}. The data acquired from a Na$_{1-\delta}$Fe$_{1-x}$Rh$_x$As sample with $x=0.019$, is presented in the Fig.~\ref{fig-ARPES2} and Fig.~\ref{fig-ARPES}. In Fig.~\ref{fig-ARPES2} the energy distribution curve (EDC) as well as the momentum distribution curve (MDC) are shown. From this measurement one can infer the good crystal quality because the upper bounds for the energy distribution at 15~K is 5~meV. Additionally, possible misalignments in crystal portions within the probed spot $(0.1\times0.1\,\mathrm{mm})$ are smaller than 1° (cf. Fig.~\ref{fig-ARPES2}~(b)) demonstrating a low mosaicity of our crystals and, hence, their good quality. Regarding the doping with Rh 4$d$ orbitals, the ARPES spectra taken throughout the Brillouin zone show that all detected features are usual for the electronic structure of iron-based superconductors. Therefore the anticipated enhanced hybridization of larger 4$d$ orbitals with As $p$ states does not significantly alter the electronic structure of this material.

Fig.~\ref{fig-ARPES}~(a) shows the evolution of the energy-momentum cut, passing through the $\Gamma$ point with increasing temperature. A difference between the three spectra recorded at higher temperatures, 19.1-22.4~K, and the two spectra recorded at 16.3-17.7~K can already be noticed in the raw data: the lowest-energy appear sharper at lower temperatures. Comparison of energy distribution curves (EDCs) for a range of temperatures confirms the presence of a superconducting transition, presented in Fig.~\ref{fig-ARPES}~(b). Three spectra recorded at temperatures from
19.1 to 22.4~K are virtually the same, while the difference occurring between 16.3 and 19.1~K is obvious - upon cooling, the coherence peak becomes more pronounced and the leading edge of the spectrum is shifted towards higher binding energies - implying the observation of the superconducting transition with a critical temperature of 18-19~K.

\subsection{Phase diagram}

\begin{figure}
\centering
\includegraphics[width=8.5cm]{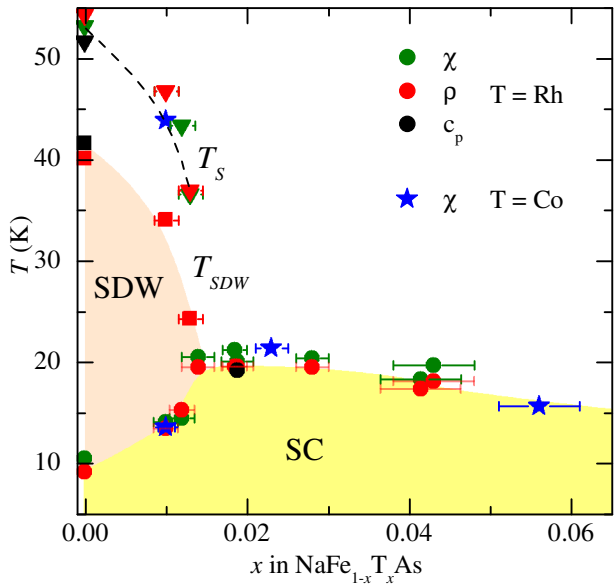}
\caption{(Color online) Electronic phase diagram of Na$_{1-\delta}$Fe$_{1-x}$Rh$_x$As inferred from magnetization, resistivity and specific heat measurements. Critical temperatures of Na$_{1-\delta}$Fe$_{1-x}$Co$_x$As are added to show the generic behavior upon electron doping in Na$_{1-\delta}$FeAs.}
\label{fig-10}
\end{figure}

Summarizing the data on $T_c$, $T_{SDW}$, and $T_S$ obtained from the magnetization, resistivity and specific heat measurements for the Na$_{1-\delta}$Fe$_{1-x}$Rh$_x$As single crystal series we construct an electronic phase diagram (see Fig.~\ref{fig-10}). The superconducting transition temperature in the Rh series is enhanced from 10~K to 21.9~K for the optimally Rh-doped sample ($x = 0.019$), and decreases for higher doping levels. At $x=0.018$ the typical anomalies in the measured physical properties associated with the phase transition are completely suppressed which seems to point to a fully suppressed SDW formation. Note, that for Co-doped  Na$_{1-\delta}$FeAs optimum $T_c$ and a full SDW suppression have been reported at slightly higher doping levels\cite{Wang2012,Wright2012}, but could be brought in accordance with optimal Rh-doped crystals in the range of the error bars of the EDX measurement. For further comparison, we include our data for Na$_{1-\delta}$Fe$_{1-x}$Co$_x$As with $x$~=~0.01, 0.023 and 0.056. The most striking observation is that both phase diagrams are obviously very similar, implying that both dopants have a very similar impact on the electronic structure.


If we now take the analogous findings for Rh- and Co-doped BaFe$_2$As$_2$\cite{Ni2009} into account, our data imply a generic impact of Rh and Co doping in both the BaFe$_2$As$_2$ and NaFeAs superconductors. Accordingly, the structural and SDW transition temperatures as well as $T_c$ depend primarily on the doping level, irrespective on the actual dopant. This suggests that potential steric effects due to larger atomic radii of the different dopants are without strong influence on the phase diagram. In contrast, in the highly pressure sensitive CaFe$_2$As$_2$ steric effects do play an important role upon Co and Rh doping \cite{Ran2014}.


\section{Summary and Conclusion}

We have performed structural, magnetic, transport, specific heat and ARPES measurements on Na$_{1-\delta}$Fe$_{1-x}$T$_x$As single crystals with
T~=~Co, Rh. A temperature-composition phase diagram is constructed based on these data. The typical suppression of the structural and magnetic phase upon electron doping is observed as well as the typical superconducting dome having its maximum near the complete suppression of the magnetic phase.

For the Na$_{1-\delta}$Fe$_{1-x}$Rh$_x$As series, it is shown that the superconducting transition temperature ($T_c$) is enhanced from 10~K to 21.9~K in the optimally doped sample while the volume fraction increases to 100\%. The SDW transition anomalies observed for low dopant concentrations are fully suppressed at a Rh content equal to $x=0.018$ which is somewhat less than that for the Co-doped pendant\cite{Wang2012,Wright2012}.

We found that the Co- and Rh-doped Na$_{1-\delta}$Fe$_{1-x}$T$_x$As yield the same electronic phase diagram for formal electron doping up to $x\sim0.05$.  Thus, we conclude that Co as well as Rh doping are of generic nature in many FeAs superconductors. A significant variation of the electronic structure due to the spatially more extended 4$d$ orbitals are not observed experimentally supporting our generic picture.

\begin{acknowledgments}

The authors thank M. Deutschmann, S. Pichl, C. Malbrich, K. Leger and S. Gass for technical support, L. Giebeler for assistance with the XRD measurements and A. Voss for the chemical analysis. 

This work has been supported by the Deutsche Forschungsgemeinschaft through the Priority Programme SPP1458 (Grant No. BE1749/13 and grant No. BU887/15-1), and through the Emmy Noether Programme in project WU595/3-1 and  WU595/3-2 (S.W.). 
Financial support by RFBR 12-03-91674-ERA\_a, 12-03-01143\_a and 12-03-31717 (M.R.) is cordially acknowledged. S.W. thanks the BMBF for support in the frame of the ERA.Net RUS project (project 01DJ12096, FeSuCo). S.B. thanks for support by BO1912/2-2 and BO1912/3-1.

\end{acknowledgments}

\bibliographystyle{apsrev4-1}
%

\end{document}